\documentclass[preprint,pra,aps,showpacs,preprintnumbers,amsmath,amssymb,eqsecnum]{revtex4}

\usepackage{graphicx}
\usepackage{dcolumn}
\usepackage{bm}
\usepackage{verbatim}
\usepackage{epsfig}
\usepackage{epstopdf}

\newcommand\beq{\begin{equation}}
\newcommand\eeq{\end{equation}}
\newcommand\bea{\begin{eqnarray}}
\newcommand\eea{\end{eqnarray}}

\def\a{{\alpha}}

\def\s{{s}}

\def\half{\frac {1} {2}}

\def\x0{{{\bf x}_0}}

\def\up{{\uparrow}}
\def\down{{\downarrow}}

\begin{document}



\title{Incompatible Multiple Consistent Sets of Histories and Measures of Quantumness}

\author{J.J.Halliwell}%

\affiliation{Blackett Laboratory \\ Imperial College \\ London SW7
2BZ \\ UK }



\begin{abstract}
In the consistent histories (CH) approach to quantum theory probabilities are assigned to histories
subject to a consistency condition of negligible interference.
The approach has the feature that a given physical situation admits multiple sets of consistent histories that cannot in general be united into a single consistent set, leading to a number of counter-intuitive or contrary properties if propositions from different consistent sets are combined indiscriminately.
An alternative viewpoint is proposed in which multiple consistent sets are classified
according to whether or not there exists 
{\it any} unifying probability for combinations of incompatible sets which replicates the consistent histories result when restricted to a single consistent set.
A number of examples are exhibited in which this classification can be made, in some cases with the assistance of
the Bell, CHSH or Leggett-Garg inequalities together with Fine's theorem.
When a unifying probability exists logical deductions in different consistent sets can in fact be combined, an extension of the ``single framework rule''. 
It is argued that this classification coincides with intuitive notions of the boundary between classical and quantum regimes and in particular, the absence of a unifying probability for certain combinations of consistent sets is regarded as a measure of the ``quantumness'' of the system. 
The proposed approach and results are closely related to recent work on the classification of quasi-probabilities and this connection is discussed. 
\end{abstract}









\maketitle

\section{Introduction}

A considerable amount of contemporary theoretical and experimental research is devoted to elucidating the counter-intuitive nature of quantum-mechanical phenomena. Ever since the birth of quantum theory, features which defy classical explanation have continued to fascinate \cite{QM}. At the same time, a parallel programme has concerned itself with what is perhaps the opposite issue, which is to explain the emergence of a quasi-classical domain from an underlying quantum description \cite{CM1,CM2,CM3}.

The phenomenon of ``quantumness'' can be characterized in many ways but it is typically linked with, for example, interferences, the breakdown of classical logic, entanglement and violation of the Bell inequalities. Likewise classicality is defined in numerous ways but it is linked with decoherence and the assignment of probabilities which indicate correlations in time according to classical equations of motion.
However, the typical definitions of quantumness and classicality are quite far apart. The quasi-classical realm is often depicted as an asymptotic regime described by very coarse grained variables suffering negligible interference \cite{CM1,CM2,CM3}. At the other end of the scale there are situations which on the face of it appear to be quantum-mechanical in nature but can be modeled in classical terms, so sit very close to the classical-quantum boundary.

The consistent histories (CH) formulation of quantum theory was first formulated over thirty years ago and continues to be a source of interest and useful applications
\cite{CM1,GH1,GH2,GH3,Gri1,Gri2,Gri3,GriEPR,Omn1,Omn2,Omn3,Hal2,Hal3,Hal4,Hal0,DoHa,DoK,Ish,IshLin,Dio,Cra}.
It was formulated in order to free standard quantum theory from dependence on an assumed separate classical domain, as is required to extend quantum theory to the whole universe, i.e. to quantum cosmology \cite{HalQC}, since there can be no separate classical domain in the very early universe.
Instead of a classical domain the approach focuses on finding the situations in which probabilities may be assigned to histories and hence, to which classical logic may be applied.
This framework has turned out to be a very useful one for studying the emergence of classical behaviour from quantum theory.
One can also examine from this framework many of the so-called paradoxes of quantum theory, some of which are then seen to arise from indiscriminate use of classical logic.
Furthermore, the focus on histories of the system, rather than events at a single time, means that the approach naturally adapts to situations in which time enters in a non-trivial way, or indeed in which time is entirely absent, as is the case in quantum cosmology. In all its applications it would probably reasonable to say that the CH approach has enjoyed considerable success. However, despite these successes certain aspects of the CH approach have met with resistance.

The initial mathematical aim of the approach is, for a system in a given initial state $\rho$ , to determine which sets of histories, characterized by time-ordered strings of projection operators $P_{a_n} (t_n) \cdots P_{a_1} (t_1) $ (or sums of such strings), have negligible interference and therefore, to which probabilities of the form
\beq
p(a_1, a_2, \cdots a_n) = {\rm Tr} \left( P_{a_n} (t_n) \cdots P_{a_1} (t_1) \rho P_{a_1} (t_1) \cdots P_{a_n} (t_n) \right),
\label{prob}
\eeq
may be assigned which obey all the usual sum rules. Such sets histories are then said to be consistent.
The sequences of alternatives described by those histories may then be discussed using the rules of classical logic. One can then, for example, address whether the correlations these probabilities indicate are well-approximated by classical dynamical laws.

The procedure, however, has a particular feature which is perhaps the greatest source of criticism. This is that a given physical situation defined by a fixed initial state (and in some cases a fixed final state) in general admits {\it more than one} consistent set of histories which are incompatible, i.e. cannot be combined into a larger, single consistent set.
Furthermore, in situations where there is a fixed final state, it is easy to find examples where two non-commuting observables, such as spin in two different directions, each have probability $1$ in different incompatible consistent sets, at variance with naive notions of the uncertainty principle. Perhaps even more challenging, there exist examples with ``contrary'' properties,
in which a certain variable has probability $1$ in one consistent set and an orthogonal variable has probability $1$ in a different consistent set.

These intuitively challenging features clearly mean that it is not possible in general to
associate definite values with consistent sets of histories, and indeed the CH approach is not, and was not intended to be, a hidden variables theory. Nevertheless
these aspects of the CH approach have led some to question the utility of the entire approach or to suggest modifications or additional conditions which might restrict some of the more challenging examples \cite{CHcrit,DoK,Kent,Wall}.
Some of these features were first noticed by Griffiths in the very first paper on the subject \cite{Gri1} and he has since offered numerous robust defences of the criticisms \cite{GriC,GrHa}. In brief, he argues that one of the rules of the game is that any logical deductions must be made within the framework of single consistent set of histories -- one cannot combine incompatible sets. If one accepts this ``single framework rule'', the intuitively challenging features indicated above are ruled out and in particular explicit logical contradictions are not possible.

Although Griffiths' procedure for handling multiple consistent sets is very reasonable in operational terms, there are some situations in which we may have good physical reasons for wishing to talk about properties living in incompatible consistent sets but the properties of multiple consistent sets outlined above create an obstruction. This issue arises in particular 
if we attempt to use the CH approach to delineate a clear boundary between the classical and quantum regimes. The point is that in characterizing the classical regime, we would like to be able to talk about complementary quantities, such as positions and momenta, in a single logical framework so that we could discuss the logical connections between them. However, the non-commutativity of these quantities means they are not in general found in a single consistent set -- they are usually found only in different incompatible sets. Hence in this sort of situation it would be extremely valuable to determine if, for at least some physically interesting examples, we can in fact combine certain types of incompatible consistent sets in some way, i.e. to see if there is any way around the single framework rule.

The purpose of this paper is to propose an alternative viewpoint on the use and interpretation of multiple consistent sets which 
extends the single framework rule and
helps to characterize the classical-quantum boundary in a way that meets intuitive expectations.
The proposed approach is to relax the focus on the standard formula for probabilities for histories used in the CH approach Eq.(\ref{prob})
and instead ask, in each situation where incompatible consistent sets exist, if there is {\it any} unifying probability for some of the combined incompatible sets which replicates the consistent histories result Eq.(\ref{prob}) when restricted to a single consistent set.
Although it is clearly not possible in general to find such a unifying probability we shall show that there
are many examples of multiple consistent sets in which such a probability (in general non-unique) does in fact exist
and it is then legitimate to combine logical statements from different consistent sets.
In simple examples, we can use the Bell \cite{Bell} and CHSH \cite{CHSH} inequalities together with Fine's theorem \cite{Fine1,Fine2} to make this classification.
This procedure leads to a natural classification of multiple consistent sets, which is physically motivated and in particular meets the desired objective of characterizing the classical-quantum boundary.
Furthermore, from this perspective, the existence of  multiple consistent sets without a unifying probability is then simply a measure of the ``quantumness'' of the system. It is not, as some have suggested, a problem with the consistent histories approach.

There have been a number of earlier proposals to classify consistent sets of histories, most notably by Kent \cite{Kent} and Wallden \cite{Wall}.
They were mainly motivated by a desire to eliminate the most challenging examples of multiple consistent sets, namely those with contrary properties mentioned above. 
However, as we shall see, this classification still allows multiple consistent sets that contain some significant quantum behaviour. The focus of the present attempt, by contrast, is to seek a physically motivated classification more in line with our intuitive understanding of classical and quantum.

Some authors regard such classifications as ``set selection principles" which inform the interpretation of the formalism. In particular, it is sometimes asserted that, ``nature somehow chooses one set of histories from among those allowed, and then randomly chooses to realize one history from that set" \cite{Kent}. Here, no claims are being put forwards about whether particular histories or sets of histories are realized and it is not the aim to find a set selection principle that will complete the programme sketched in, for example Ref.\cite{DoK}.
Rather the main aim is to explore the consequences of extending the single framework rule and determine how the intuitively understood classical-quantum boundary is expressed through the consistent histories approach and in particular how it relates to the properties of multiple consistent sets.


We summarize the key mathematical properties of the consistent histories approach in Section 2. Multiple consistent sets and the proposed alternative approach to handling them are discussed in Section 3. A number of examples with a unifying probability are given in Section 4, along with a brief discussion of the possible consequences of the non-uniqueness of the unifying probability.
Examples without a unifying probability are given in Section 5. The relationship to set selection principles of Kent and of Wallden are discussed in Section 6. A particularly important example of multiple consistent sets, concerning the question as to whether quasi-classical behaviour persists to the future, is discussed in Section 7. The relationship between this work and a recent work on the classification of quasi-probabilities is described in Section 8. We summarize and conclude in Section 9.

\section{The Consistent Histories Approach}

We briefly review the formalism of the consistent histories approach. Full details may be found in many different places
\cite{CM1,GH1,GH2,GH3,Gri1,Gri2,Gri3,GriEPR,Omn1,Omn2,Omn3,Hal2,Hal3,Hal4,Hal0,DoHa,DoK,Ish,IshLin,Dio,Cra}.
Alternatives at a single moment of time are
represented by a set of projection operators $\{ P_a \}$,
satisfying the conditions
\bea
\sum_a P_a &=& 1,
\label{PDI}
\\
P_a P_b &=& \delta_{ab} P_a,
\eea
where we take $a$ to run over some finite range.
A (homogenous) history is represented by a time-ordered string of projections,
\beq
C_{\a} = P_{a_n} (t_n) \cdots P_{a_1} (t_1),
\label{2.3}
\eeq
where $C_{\a}$  is usually referred to as a class operator. One may also consider class operators defined by sums of strings (and these are known as inhomogeneous histories).
Here the projections are in the Heisenberg picture and $ \a $ denotes
the string $ (a_1, \cdots a_n)$.
The class operator Eq.(\ref{2.3}) satisfies the conditions
\beq
\sum_{\a} C_{\a} = 1,
\label{2.4}
\eeq
and also
\beq
\sum_{\a} C^{\dag}_{\a} C_{\a} = 1.
\label{2.5}
\eeq
Probabilities are assigned to histories via the formula
\beq
p(\a) = {\rm Tr} \left( C_{\a} \rho C_{\a}^{\dag} \right),
\label{2.5}
\eeq
where $\rho$ is the initial density operator.
These probabilities are clearly positive and normalized
\beq
\sum_{\a} p(\a ) = 1,
\label{2.6}
\eeq
which follows from Eq.(\ref{2.5}).

The sample space for a quantum system consists of a projective decomposition of the identity \cite{Gri2,Gri3}.
Hence, for alternatives at a single moment of time the probabilities $p(a)$ are defined on the sample space consisting of the projective decomposition of the identity Eq.(\ref{PDI}). For the case considered here in which there are $n$ sequential non-commuting projectors, the corresponding decomposition of the identity Eq.(\ref{2.4}) is not a projective one. However, it may be made so using the temporal logic approach of Isham et al \cite{Ish,IshLin}, in which the $C_{\a}$ is replaced by an $n$-fold tensor product of projectors acting on an $n$-fold tensor product Hilbert space. This then is a projective decomposition of the identity, on the larger Hilbert space, and
defines the {\it histories sample space} for the probabilities Eq.(\ref{2.5}) \cite{Gri2,Gri3}.


This assignment of probabilities to non-commuting quantities such as those appearing here
is only meaningful if there is no interference between pairs of histories and this is measured by the decoherence functional,
\beq
D(\a, \a') = {\rm Tr} \left( C_{\a} \rho C_{\a'}^{\dag} \right).
\eeq
It satisfies the conditions
\bea
D(\a, \a') &=& D^* (\a', \a),
\\
\sum_{\a} \sum_{\a'} D(\a, \a') &=& 1,
\label{2.10}
\eea
and note that the probabilities are given by its diagonal elements
\beq
p(\a) = D(\a, \a).
\eeq
The simplest and most important condition normally
imposed is that the probabilities should satisfy the probability sum rules and this is the case if and only if
\beq
{\rm Re} D(\a, \a') = 0, \ \ \ \a  \ne \a',
\label{2.13}
\eeq
for all pairs of histories $\a, \a'$, a condition is referred to as
consistency of histories.
In many practical situations, there is present a physical mechanism
(such as coupling to an environment) which causes Eq.(\ref{2.13})
to be satisfied, at least approximately, and in such situations, it is
typically observed that the imaginary part of the off-diagonal
terms of $ D(\a, \a')$ vanish as well as the real part.
It is therefore of interest to consider
the stronger condition of decoherence, which is
\beq
D(\a, \a') = 0, \ \ \ \a \ne \a'.
\label{deco}
\eeq
This stronger condition is related to the existence of records \cite{GH2,Hal4}.

In the search for further conditions for the assignment of probabilities, it is useful to consider the quasi-probability
\beq
q(\a) = {\rm Re} {\rm Tr} \left( C_{\a} \rho \right).
\label{2.17}
\eeq
Because it is linear in the $C_{\a}$, this quantity sums to $1$ and
also satisfies the
probability sum rules, but it is not in general positive.
However, it is closely related to the probabilities Eq.(\ref{2.6}), because
Eq.(\ref{2.4}) implies that
\bea
q (\a ) &=& {\rm Tr} \left( C_{\a} \rho C_{\a}^{\dag} \right) +2  {\rm Re} {\rm Tr} \left( C_{\a} \rho
{\bar C}_{\a}^{\dag} \right),
\nonumber \\
&=& p(\a) + 2 {\rm Re} D(\a, {\bar \a}).
\label{2.18}
\eea
Here $ {\bar C}_{\a} $ denotes the negation of the history $C_{\a}$,
\beq
{\bar C}_{\a} =1 - C_{\a}.
\label{2.19}
\eeq
This means that when there is consistency the probabilities
are given by the simpler expression
\beq
p(\a ) = q (\a).
\label{2.20}
\eeq
Consistency therefore ensures that $q(\a)$ is real and positive, even though it
is not in general.

These properties suggest an alternative to the consistent histories approach in which
the
probabilities are given by $q(\a)$, subject only to the requirement
that
\beq
q( \a) \ge 0,
\label{2.22}
\eeq
a condition referred to as {\it linear positivity} \cite{GoPa}. (The sample space is still the histories sample space described above).
These probabilities agree with the usual assignments $p(\a)$ when there is consistency,
but this condition is clearly weaker than consistency so the reverse is not true.

These properties also suggest an alternative condition, named partial decoherence \cite{Hal0}, which is the requirement that the probabilities satisfy Eq.(\ref{2.20}), or equivalently, that each history has zero interference with its negation. This condition is stronger than linear positivity, weaker than decoherence, but can be weaker or stronger than consistency.

The above formulae easily generalize to the case in which there is a final state $\rho_f$, as is the case in post-selection. The probability for histories then is
\beq
p ( \a)  = \frac {1} { {\rm Tr} (\rho_f \rho) } {\rm Tr} \left( \rho_f C_{\a} \rho C_{\a}^{\dag} \right),
\label{probf}
\eeq
(and similarly for the decoherence functional)
and the quasi-probability Eq.(\ref{2.17}) generalizes to
\beq
q(\a) =  \frac {1} { {\rm Tr} (\rho_f \rho) } {\rm Re} {\rm Tr} \left( \rho_f C_{\a} \rho \right).
\label{quasif}
\eeq

Both consistency and linear positivity suffer from some unusual properties under subsystem composition (also known as the Dio\'si test \cite{Dio}). This is the requirement that the condition for probability assignment for a composite system consisting of two uncorrelated and non-interacting parts $A$ and $B$ should be equivalent to the probability assignments for $A$ and $B$ separately. Partial decoherence comes very close to meeting subsystem composition, but narrowly fails for inhomogeneous histories. Only decoherence, Eq.(\ref{deco}), satisfies it exactly (although it can also be satisfied by the very weak procedure outlined below).

Even weaker conditions ensuring the assignment of probabilities to certain types of histories are possible in some situations. The above formulae are all concerned with the conditions under which a specific formula for the probabilities for histories may be successfully assigned. However, there are certain situations when it is of interest to ask the simpler question as to whether any probability exists, subject to certain conditions. This is closely related to the question of determining whether certain situations admit a local hidden variable description.

To give a specific example,
suppose we have a system such as a spin system with dichotomic variables in which there are three possible projections and for which there exist non-negative probabilities $p(a_1,a_2)$, $p(a_2,a_3)$ and $p(a_1,a_3)$, non-negative either because they correspond to pairs of commuting observables, or because linear positivity is satisfied for each pair. Suppose, however, that consistency of histories and linear positivity fail to yield a formula for a positive probability $p(a_1, a_2,a_3)$ matching the three given pairwise probabilities. Does this mean that there is no probability?  The answer is that sometimes there is. In particular, {\it some} probability exists matching the three pairwise probabilities if and only if the Bell inequalities \cite{Bell} are satisfied:
\beq
-1 \le C_{12} + C_{13} + C_{23} \le 1 + 2 \ {\rm min} \{ C_{12}, C_{13}, C_{23} \},
\label{Bell}
\eeq
where the $C_{ij}$ are the correlation functions of the three probabilities, for example,
\beq
C_{12} = \sum_{a_1 a_2}  a_1 a_2  \ p(a_1, a_2).
\eeq
Similarly, when we have four
pairwise probabilities the necessary and sufficient condition for the existence of an underlying probability
is the set of eight CHSH inequalities \cite{CHSH}.
These important results (Fine's theorem \cite{Fine1,Fine2}) are simply existence theorems -- they do not provide a general formula for the probabilities of quantum-mechanical form.
Nevertheless, there is a significant difference between the case where some probability exists and the case where no probability exists.
(Note however that the problem of matching a probability to a given set of marginals is in general a very difficult problem \cite{GaMer}).

This procedure provides a way of assigning probabilities to histories or to non-commuting observables this is demonstrably weaker than linear positivity. It also has the appealing feature that it is compatible with subsystem composition \cite{HaYe}.
We will make use of this procedure in what follows to analyze multiple consistent sets.

Significantly, the sample space for probabilities defined in this way is no longer the histories sample space described above, but is instead the sample space of a local hidden variable theory which will may have the form, for example of a classical phase space \cite{Spe}.
This is a step outside the conventional CH framework but confers some useful advantages, as we shall see.


\section{Multiple Consistent Sets}

In the Copenhagen interpretation, it is usually asserted that the only quantities we can talk about in an unambiguous way are quantities that are physically measured. By contrast, in the CH approach, it is claimed that we can extend that discussion to quantities that are not measured, using consistent sets of histories and classical logic. For example, we can talk about what is going on with a quantum system between measurements, or after initial preparation but before the first measurement takes place. Or, we can talk about past histories of the universe even though the only measurements made are in the present moment.
However, this extension from measured to unmeasured quantities turns out to be subtle due to the existence of multiple consistent sets and care is required in terms of deciding what sort of logical deductions can be made.

\subsection{A Simple Spin Example}

To exemplify this, consider the following example first given by Griffiths in his original paper on the CH approach \cite{Gri1}.
The example is a simple spin system, initially in the up state in the $z$-direction $| \up \rangle$ and post-selected to be in the $|+\rangle$ state in the $x$-direction, where
\beq
|+\rangle = \frac {1} {\sqrt{2}} \left(  | \up \rangle + | \down \rangle \right).
\eeq
We take the Hamiltonian to be zero and ask what happens between initial preparation and final measurement using a projector $P_a$. The probability is given by Eq.(\ref{probf}) which turns out to be
\beq
p(a) = 2 | \langle + | P_a | \up \rangle |^2.
\eeq
If we take $P_a$ to project onto the $z$-spin we get $p_z(\up)=1$, $p_z(\down) =0$, so the two histories are consistent and have probability $1$ for spin up. On the basis of this we might be inclined to say the spin is up in the $z$-direction between measurements. However, if we take $P_a$ to project onto the $x$-spin we get $p_x(+)=1$ and $p_x(-)=0$. So again the histories are consistent but we get probability $1$ for spin $+$ in the $x$-direction, which suggests that the spin takes definite value in the $x$-direction between measurements.

One can look at a more complicated history in which both spins are projected on intermediately, using a class operator of the form
\beq
C_{a_1 a_2} = P_{a_2}^z P_{a_1}^x.
\label{Czx}
\eeq
However, it is easily shown that such histories are not consistent. That is, we cannot combine the probabilities for the two consistent sets into a single consistent set with probability given by the formula Eq.(\ref{probf}). (One can work instead with the opposite operator ordering in the class operator Eq.(\ref{Czx}) and the resulting histories then are in fact consistent, but this is the essentially trivial case in which the first projector coincides with the initial state and the second projector coincides with the final state. We will suppose that the physical situation dictates that the above ordering is the appropriate one).

It would clearly be incompatible with the uncertainty principle to assert that both the $x$ and $z$ spin are definite in this way, so what are we to make of these properties?
As indicated in the Introduction, Griffiths argues that in any application of classical logic to a quantum system with consistent sets of histories, any deduction must be made within the framework of a single consistent set. Deductions belonging to different consistent sets cannot be combined. In this example, we therefore cannot deduce that the spin takes definite values in both directions. This example is the simplest example of a essentially universal feature of the CH approach which is that a given physical situation in which we attempt to ascertain what is happening between measurements admits a number of different consistent sets which, if taken at face value, appear to have properties at variance with certain intuitive notions of basic quantum mechanics.

\subsection{An Alternative Approach}

Let us now focus in more general terms on what it would mean to think of the spins in both directions as possessing definite values. The CH approach does not allow this. However, the CH approach is specifically tied to probabilities for histories given by Eq.(\ref{probf}). But what would be happen if we worked instead with other probabilities, such as the quasi-probability Eq.(\ref{quasif}), or the more general approach of the last section not involving a specific formula? It is easy in this simple example to write down a unifying probability that does the job, namely
\beq
p(a_1, a_2) = p_x (a_1) p_z (a_2).
\eeq
This is clearly positive and also matches the above marginals for $p(a_1)$ in the $x$-direction and $p(a_2)$ in the $z$-direction. In particular, $p(+,\uparrow)=1$ and the remaining components are zero.
Therefore, a consistent joint probability for both spins exists, although it is not the CH probability, Eq.(\ref{probf}).
(One could also try the quasi-probability Eq.(\ref{quasif}) associated with the class operator Eq.(\ref{Czx}) but this has a negative component).

Because {\it some} probability for both spins exists, there is no contradiction in asserting that both spins take definite values. Indeed, it is well-known that all the predictions of a single spin system of this type may be replicated by a hidden variable theory \cite{Bell}.

These important observations lead to the following strategy. In each situation in which there are two or more consistent sets, we can ask if there is {\it any} unifying probability for the combined consistent sets
which matches the CH probabilities when restricted to each individual consistent set. If such a probability exists, then we can assert that logical statements from different consistent sets can be combined. (Here, we are of course invoking the well-known connection between Boolean logic and probability emphasized by Omn\`es \cite{Omn1,Omn2,Omn3}).

This approach clearly steps beyond the conventional CH approach although it is not in contradiction with it. In the CH approach with the single framework rule statements from different consistent sets cannot be combined, but here we argue that they can be, without contradiction, in some, but not all circumstances. It thus seems reasonable to introduce an {\it extended single framework rule}:

\begin{itemize}

\item
{ Logical statements from different consistent sets cannot be combined unless a unifying probability for those consistent sets exists.}

\end{itemize}

Note that a given system may have many incompatible consistent sets and generally, only some of those sets can be combined in the way described above. This means that we still cannot, in general, assign definite values to all the quantities describing the system. Although we may do so within the framework of those sets for which there is a unifying probability in which we are contemplating measuring some of the variables and then using the probability to deduce the values of other unmeasured quantities lying within the unified consistent sets.

\def\CS{\rm CS}

With three or more consistent sets one can encounter more complicated combinations of incompatible sets but
the extended single framework rule continues to apply.
Consider for example a situation in which there are three consistent sets, which we denote $\CS_1$, $\CS_2$, $\CS_3$ and suppose that there exists a unifying probability for two of the possible pairs, $\CS_1, \CS_2$ and $\CS_2, \CS_3$, but not for the pair
$\CS_1, \CS_3$. This means first of all that there is no unifying probability for all three sets. Secondly, it 
means that we are allowed to make logical deductions within the unified sets $\CS_1 \cup \CS_2$ and $\CS_2 \cup \CS_3$, but the extended single framework rule means that we are not allowed to combine logical statements between these two unified sets, and in particular we cannot invoke a ``transitivity'' argument involving $\CS_2$ to combine statements between $\CS_1$ and $\CS_3$.

Note also that since the probabilities for each consistent set may be expressed, via Eq.(\ref{2.20}),  in terms of the Goldstein-Page quasi-probability, Eqs.(\ref{2.17}), it is {\it always} possible to write down a unifying quasi-probability for the combined multiple consistent sets, namely, the quasi-probability Eq.(\ref{2.17}) obtained by combining the class operators from each set. This is clearly a natural thing to check but it may or may not be positive in each case. 
If it is not, there often exist other ways of constructing a unifying probability as outlined in the last section.

As stated, the general search for a unifying probability outlined in the last section entails a switch from the histories sample space to a local hidden variable theory sample space. This is a significant change but carries two key advantages.
First of all, it addresses the ontological questions surrounding multiple consistent sets (i.e. to what extent can we assign definite values to the histories in different consistent sets), in a way that is thoroughly consistent with conventional thinking around hidden variable theories. Secondly, it also brings the CH approach (with the extended single framework rule) into a position where the quantum-classical boundary is characterized in a way closer to intuition and with other standard definitions of that boundary. We will see this in detail in the examples of the following sections.

\subsection{Another Simple Example}

Another simple but very different example, given by Omn\`es \cite{Omn2}, consists of a free particle in three dimensions initially in an outgoing spherical wave state $ | \psi \rangle $ (e.g. a radioactive decay) and the final state $\rho_f$ is a measurement which localizes in position. We can then ask what happens between initial and final states at a sequence of times $t_1, t_2, \cdots t_n $.
There are a number of different consistent sets. First, we could take the class operators to describe a sequence of coarse-grained projections onto ranges of positions denoted by centres 
${\bf x_1}, {\bf  x_2} \cdots  {\bf x_n}$, not too closely spaced in time and onto reasonably large spatial regions. These histories have probability close to $1$ when the spatial regions lie along a straight line path and are approximately zero otherwise, hence are consistent histories. Secondly, we could instead do a sequence of projections onto ranges of momenta with centres 
${\bf p_1}, {\bf  p_2}, \cdots { \bf p_n}$ and we will find the probability is close to $1$ for momenta close to the expected classical trajectory.

One cannot in general combine these different sets into a single consistent set since they refer to incompatible quantum properties. However, it is easy to see that there is a unifying probability for these two incompatible sets, namely the simple product,
\beq
p ({\bf x_1}, {\bf  x_2} \cdots  {\bf x_n} ) \ p({\bf p_1}, {\bf  p_2}, \cdots { \bf p_n}),
\label{3.5}
\eeq
which trivially matches the desired marginals. Here the sample space is the Cartesian product of $n$ (discretized) classical phase spaces for the point particle in three dimensions.

A third possibility is to consider also
the projections onto the state or its negation at each time, using a projector $P_a$, with $a=1,2$, where $P_1 = | \psi \rangle \langle \psi | $ and $ P_2 = 1 - P_1 $, and we denote these histories by $a_1, a_2, \cdots a_n$.
These histories will be exactly consistent with probability $1$ for the single history consisting of the evolving state and probability $0$ for any other history. There is then the possibility of combing all three of the above consistent sets using the probability
\beq
p ({\bf x_1}, {\bf  x_2} \cdots  {\bf x_n} ) \ p({\bf p_1}, {\bf  p_2}, \cdots { \bf p_n})
\  p( a_1, a_2 \cdots a_n),
\eeq
which again matches the desired marginals. The sample space is then the Cartesian product of the sample space for Eq.(3.5) with the histories sample space for $p(a_1, a_2, \cdots a_n)$.

In both of the examples in this section, the desired unifying probability is easily obtained by taking a product of the probabilities for each consistent set (although this is not necesssarily the only way to obtain it). Hence these examples are reasonably trivial, because each consistent set has no alternatives in common with the other sets. The more challenging (and perhaps more common) case is that in which the consistent sets have partial overlap and this we now consider.

\section{Examples with a Unifying Probability}

\subsection{The EPRB State}

We first consider a particularly instructive example in which there is a unifying probability for some parameter ranges but not for others. The example is the standard EPRB situation,
in which
we consider a pair of particles $A$ and $B$ whose spins are in the singlet state,
\beq
| \Psi \rangle = \frac {1} {\sqrt{2}} \left( | \uparrow \rangle \otimes | \downarrow \rangle
- | \downarrow \rangle \otimes | \uparrow \rangle \right),
\eeq
where $ |\uparrow \rangle $ denotes spin up in the $z$-direction.  We consider the spins of
particle $A$ in the directions characterized by unit vectors ${\bf a}_1$ and
${\bf a}_2$ and with values $s_1$, $s_2$;
and on particle $B$ in directions ${\bf a}_3$ and $ {\bf a}_4$ with values $s_3$, $s_4$, where each $s$ may take values $\pm 1$.
The probabilities for pairs of such alternatives, one on $A$, one on $B$ are each of the
form
\beq
p(s_1, s_3) = {\rm Tr} \left( P_{s_1}^{{\bf a}_1} \otimes P_{s_3}^{{\bf a}_3} | \Psi \rangle
\langle \Psi | \right),
\eeq
where the projection operators are given by
\beq
P_{s}^{\bf a} = \half \left( 1 + s {\bf a} \cdot \sigma \right),
\eeq
where $\sigma_i$ denotes the Pauli spin matrices.  We similarly define three more pairwise probabilities $p(s_1,s_4)$, $p(s_2,s_3)$, $p(s_2,s_4)$. Each of these four probabilities defines a set of ``histories" which is trivially decoherent, since the projection operators within each set commute.

Combining the above sets into larger consistent sets is non-trivial. Suppose we consider histories involving both spins of each particle.
To analyze these histories, we need the decoherence functional,
\beq
D (s_1, s_2, s_3, s_4 | s_1', s_2', s_3', s_4') =   {\rm Tr} \left( C_{s_1 s_2 s_3 s_4} | \Psi \rangle \langle \Psi |
 C^\dag_{s_1' s_2' s_3' s_4'} \right),
\label{EPRBdf}
\eeq
where
\beq
C_{s_1 s_2 s_3 s_4} = P_{\s_2}^{{\bf a}_2} P_{\s_1}^{{\bf a}_1} \otimes P_{\s_4}^{{\bf a}_4} P_{\s_3}^{{\bf a}_3}
\eeq
and we have selected an ordering in which ${\bf a}_1$ precedes ${\bf a}_2$ and ${\bf a}_3$ precedes ${\bf a}_4$. The decoherence functional is trivially zero for $s_2 \ne s_2' $ and $s_4 \ne s_4'$, but is not in general  diagonal (or diagonal in its real part) due to the presence of non-commuting operators. This means that in general the four consistent sets defined above cannot be combined into a single consistent set in which all four spin components are specified. Hence the four consistent sets are incompatible in general.

However, we can now ask whether there is a unifying probability matching the four probabilities
$ p(s_1, s_3)$, $p(s_1,s_4)$, $p(s_2,s_3)$, $p(s_2,s_4)$ from the four consistent sets. One possible way to approach this might be to try the quasi-probability
\beq
q (\s_1, \s_2, \s_3, \s_3 ) =  {\rm Re} {\rm Tr} \left( P_{\s_2}^{{\bf a}_2} P_{\s_1}^{{\bf a}_1} \otimes P_{\s_4}^{{\bf a}_4} P_{\s_3}^{{\bf a}_3} | \Psi \rangle \langle \Psi | \right),
\label{quasiEPRB}
\eeq
which clearly matches the four probabilities. It is not positive in general but will be positive for a parameter range that is larger than that for which the decoherence functional is diagonal, since it only requires the interference terms to be suitably bounded, not zero. (Hence the problem of multiple consistent sets is generally weaker for the linear positivity approach \cite{GoPa}).

But we can also ask very generally, without appealing to a specific formula, is there any probability, perhaps defined on a hidden variables sample space, which matches the four pairwise ones? The answer to this question, as indicated in Section 2, is given by Fine's theorem \cite{Fine1,Fine2}, which states that there exists a non-negative probability $p(s_1,s_2,s_3,s_4)$ matching the given four pairwise probabilities if and only if the eight CHSH inequalities hold \cite{CHSH}. These inequalities have the form
\beq
\left| C_{13} + C_{14} + C_{23} - C_{24} \right| \le 2,
\eeq
plus three more similar relations with the minus sign in the other three possible locations. Hence there is a unifying probability for the incompatible consistent sets as long as the CHSH inequalities hold.  As noted in Ref.\cite{HaYe}, these inequalities can hold even when Eq.(\ref{quasiEPRB}) is not positive.

It is instructive to consider a specific example, namely that in which measurements are made only in the $x$ or $z$ direction, so we take ${\bf a}_1 = {\bf a}_3 = (0,0,1)$  and ${\bf a}_2 = {\bf a}_4= (1,0,0)$.
The probabilities $p(s_1,s_3)$ and $p(s_2,s_4)$ then show perfect anticorrelation. It is then tempting to assert that the spins take definite values and the standard argument, essentially that of EPR \cite{EPR},  then appears to indicate that one can deduce the spin in both directions of both particles using this anticorrelation.

In a consistent histories analysis, the histories in which the spins in both directions of both particle is specified are not consistent (as is easily shown using Eq.(\ref{EPRBdf})). The anticorrelation exists within certain consistent sets, but the single framework rule forbids such logical deductions from being combined with statements made in other, incompatible sets. (See also the CH analysis by Griffiths of this situation \cite{GriEPR}).

However, from the point of view of the approach of this paper, this is not the end of the road. The correlation functions for this situation are $C_{13} = C_{23} = -1$ and $C_{14} = C_{23} = 0$ so all the CHSH inequalities are satisfied. This means that there is in fact some probability distribution coinciding with the four marginals. In general one would expect there to be a family of such distributions but it turns out in this case that the quasi-probability Eq.(\ref{quasiEPRB}) is positive so does the job. Using the explicit formula given in Ref.\cite{HaYe}, this turns out to be
\beq
q(s_1,s_2,s_3,s_4) = \frac{1}{16} (1 - s_1 s_3) (1-s_2 s_4),
\label{UPq}
\eeq
which is clearly non-negative and exhibits the desired correlations. (Since the quasi-probability turns out to be positive in this case, the sample space is in fact the usual histories sample space).

According to the extended single framework rule, logical statements from different consistent sets may be combined in this case, since a unifying probability exists. This means it is consistent to assert that the two particles have definite spins in both the $x$ and $z$ directions. Physically, it corresponds to the known fact that this situation admits a local hidden variables description. The CH approach with the usual single framework rule misses this essentially classical situation.
This example illustrates particularly clearly why it is of interest to explore an extended single framework rule.

\subsection{The EPR State}

Another closely related and pertinent example is the original EPR state \cite{EPR}, which in one dimension is the two particle state
\beq
\psi (x_1, x_2) = N \exp \left( - \frac { (x_1 - x_2)^2 } { \sigma^2 } - \sigma^2 (x_1 + x_2)^2 \right),
\label{EPR}
\eeq
where the parameter $\sigma $ may be taken to be very small and is there to make the state a normalizable Gaussian, and $N$ is a normalization factor. The state is therefore tightly peaked about $x_1 = x_2$ and
in momentum space the state $ \tilde \psi (p_1, p_2)$ is tightly peaked about $p_1 = -p_2$. One can consider two different consistent sets, one in which the positions of each particle are specified, the other in which the momenta are specified. These are characterized by projections onto small ranges of position and momentum. The probabilities indicate the correlations described above. However, one cannot in general combine these two different consistent sets into a single consistent set due to the presence of non-commuting operators, so the two sets are incompatible.

We now therefore ask if there is a unifying probability in which the coordinates and momenta of both particles are specified. This was answered by Bell a long time ago \cite{BellW}. The point is that the (regularized) EPR wave function
Eq.(\ref{EPR}) is a Gaussian, which implies that its Wigner function $W(x_1,p_1,x_2,p_2)$ is non-negative \cite{Wig}, and is precisely the desired unifying probability matching the probabilities $ | \psi (x_1,x_2)|^2$ and $ | \tilde \psi (p_1,p_2)|^2$ for the two consistent sets. The sample space is the classical phase space for two particles moving in one dimension.
It is therefore consistent to assert that the coordinates and momenta of both particles take definite values.

\subsection{Histories of a Single Spin}

Another instructive example is that provided by the spin systems studied in the Leggett-Garg inequalities -- the analogue of the EPRB situation for a single particle characterized by alternatives at three or more times \cite{LeGa}. We focus on a variable $\hat Q = \sigma_z$ which evolves under Hamiltonian $H = \half \omega \sigma_x $ (where $\sigma_x, \sigma_z$ are the usual Pauli matrices). We consider the two-time probabilities
\beq
p(s_1, s_2) = {\rm Tr} \left( P_{s_2} (t_2) P_{s_1} (t_1) \rho P_{s_1} (t_1) \right),
\eeq
where
the projectors at each time are $ P = \half ( 1 - s \hat Q)$. It is easily shown that these two-time histories are not consistent in general, however, they are for the case of a maximally mixed initial state, in which case it may be shown that
\beq
p(s_1, s_2) = \half \left( 1 + s_1 s_2 C_{12} \right),
\eeq
where the correlation function $C_{12}$ is given by
\beq
C_{12} = \half \langle  \hat Q(t_2) \hat Q(t_1) + \hat Q(t_1) \hat Q (t_2) \rangle.
\eeq
With the above Hamiltonian and choice of $\hat Q$, we have $C_{12} = \cos \omega (t_2 -t_1)$.
Details may be found, for example, in Ref.\cite{HalLG}.  This is a more general example of the ``hopping model'' \cite{Hop}.

We may also consider similar two-probabilities at times $t_2,t_3$ and times $t_1,t_3$. We thus obtain three consistent sets of histories with probabilities $p(s_1,s_2)$, $p(s_2,s_3)$ and $p(s_1,s_3)$. These are incompatible since the underlying set of histories in which $Q$ is specified at all three times is inconsistent in general. However, like the Bell and CHSH case, we can ask if there is any probability matching the three marginals and the answer is again that some probability $p(s_1, s_2,s_3)$ exists if and only if the four Bell inequalities Eq.(\ref{Bell}) hold. (In this context they are referred to as the Leggett-Garg inequalities). Again we step beyond the usual quantum-mechanical history sample space to a ``classical history'' sample space, of the type one might use in a stochastic process, in which $Q$ takes definite values at three times.

\subsection{Comments on the Non-Uniqueness of the Unifying Probability}

Armed with above examples we are now in a position to address a potentially worrisome feature of the procedure used to identify a unifying probability, namely the fact that it will in general be non-unique. A natural question to ask is whether this non-uniqueness may affect the logical or probabilistic reasoning we are seeking to apply when different consistent sets are combined.

The point here is that any such reasoning is made using only the probabilities within each separate consistent set, i.e. using only the marginals, and these are uniquely defined (even though they can be matched to a non-unique family of unifying probabilities).  Hence although in the extended framework rule we are moving from a specifc formula for the probabilities to a general formula, the two formulae must match at the level of individual consistent sets.

Differently put, the question we are interested in is whether the logical or probabilistic reasoning within a given consistent set can be unambiguously combined with the logical or probabilistic reasoning within another set. E.g. if $A$ implies $B$ in one consistent set and $B$ implies $C$ in another consistent set, does this mean that $A$ implies $C$? The answer is yes if a unifying probability exists and there is no ambiguity since the marginal probabilities used in making these deductions are uniquely defined. The existence or otherwise of a unifying probability is simply a test to make sure that such deductions can be consistently made. 

Similar statements hold in examples in which the reasoning is probabilistic rather than logical. There are actually few examples of this type, although it is still necessary to be sure that the reasoning is consistent.

Interestingly, in the example at the end of Section 4(A), in which some of the marginals are zero and hence definite logical connections can be made, it turns out at the unifying probability Eq.(\ref{UPq}) is in fact unique. This is reasonably easily seen from explicit moment expansions given in Refs.\cite{Fine2,HaYe} and the detailed proofs of Fine's theorem in Ref.\cite{Fine2}. Loosely speaking,
the anticorrelations between $s_1$ and $s_3$ and between $s_2$ and $s_4$ essentially fix 
Eq.(\ref{UPq}) uniquely. One would expect this to be true for other similar examples, since if some of the marginal probabilities are zero the corresponding components of the unifying probability (which are summed to give the marginals) must also be zero, thereby imposing significant restrictions on the possible form of the unifying probability. However, no general proof of this claim is given here but this will be investigated elsewhere.

\section{Examples without a Unifying Probability}


The EPRB and Leggett-Garg examples of the previous Section clearly supply examples without a 
unifying probability distribution if the CHSH or Leggett-Garg inequalities are violated. However, a more striking and important example
of multiple consistent sets is the three box problem \cite{3box}. This is essentially equivalent to a triple slit interference experiment. It consists of a three state system with initial state
\beq
| \psi \rangle = \frac {1} {\sqrt{3}} \left( |1 \rangle + |2 \rangle + |3 \rangle \right),
\eeq
and final state
\beq
| \psi_f \rangle = \frac {1} {\sqrt{3}} \left( |1 \rangle + |2 \rangle - |3 \rangle \right),
\eeq
We consider simple histories in which there is a projection $P$ in between the initial and final state. The probability for this is given by
\beq
p = 3 \left| \langle \psi_f | P_a | \psi \rangle \right|^2,
\eeq
where we have used the fact that $ |\langle \psi_f | \psi \rangle |^2 = 1/3$. We consider two different consistent sets. In the first set there are two histories, given by projections $P_1 = |1 \rangle \langle 1 | $ and its complement $ P_{23} = |2 \rangle \langle 2| + | 3 \rangle \langle 3 |$. We easily find that
\beq
p(1) = 1, \ \ \ \ p ( 2 \ {\rm or} \ 3 ) = 0.
\eeq
In the second set, we consider $P_2 = |2 \rangle \langle 2 |$ and its complement,
$ P_{13} = |1 \rangle \langle 1| + | 3 \rangle \langle 3 |$, and we find that
\beq
p(2) = 1,  \ \ \ \ p ( 1 \ {\rm or} \ 3 ) = 0.
\eeq
On the fact of it, this appears to be a contradictory state of affairs since in one set the system is predicted to be definitely in state $1$ and in the other set the system is definitely in state $2$. As indicated already, we are not allowed to combine logical statements in different sets. Nevertheless, this is one of the most disconcerting examples of incompatible consistent sets.

One can, as in previous examples, ask whether there is a unifying probability for both sets, as there is in some previous examples. However, it is clear that the only way to find one is to allow some of the probabilities to be negative. For example, the quasi-probabilities $ p(1) = 1 = p(2) $ and $ p(3)=-1$ are consistent with the above properties. Hence there is no unifying probability in this case and we do not expect to be able to assign definite properties across multiple consistent sets.

Physically these properties are not surprising if we consider the closely related triple slit experiment. There, we have wave functions $\psi_1, \psi_2, \psi_3 $ emerging from three slits and impinging on a detector a short distance away.
The wave functions are carefully chosen so that there are some cancellations at the detector, $\psi_2 + \psi_3 = 0$ and $\psi_1 + \psi_3 = 0$. This means that detector registers nothing if we cover up slit $1$ or slit $2$ but has a non-trivial reading if we cover slits $1$ and $2$, an apparent contradiction if viewed in classical terms.

However, these results are unremarkable from the point of view of quantum mechanics, since we know that two non-trivial wave functions may be superposed in such as way as to give zero at a particular point. From this point of view, these disconcerting features are indications of quantumness. This is consistent with, and indeed a good example of, our hypothesis that quantumness may be measured by the absence of a unifying probability.

Other examples of incompatible sets lacking a unifying probability distribution and with contrary properties
are easily found such as the hopping model of Ref.\cite{Hop}, discussed at length in Ref.\cite{Wall} and the GHZ state, discussed in Ref.\cite{Gri3}.



\section{Comparsion with earlier Set Selection Principles}

A number of previous authors have proposed set selection principles designed to eliminate the sort of behaviour exhibited in the above examples. We briefly consider the proposals of Kent \cite{Kent} and Wallden \cite{Wall}. Following their nomenclature for the moment, we will refer to the probability Eq.(\ref{2.5}) as the measure $\mu ( \alpha)$ on a set of histories. The essence of the three box problem is that it is a ``zero cover'' situation in which the coarse-graining of histories with non-zero measure leads to a history with zero measure. The set selection principles of Kent and Wallden are designed to rule out this situation.

Kent accomplishes by restricting to ``ordered consistent sets'', which is
in essence the requirement that the measure behaves in a monotonic way under coarse graining \cite{Kent}. This clearly implies that it is not possible to obtain a measure zero history by coarse-graining histories with non-zero measure.

Wallden offers the slightly weaker proposal to restrict to ``preclusive consistent sets'' \cite{Wall}. These are consistent sets of histories $\{ \alpha \} $ for which there are no zero-measure coarse grainings if $\mu(\alpha) \ne  0$.

These two principles successfully isolate the contrary features of the three box problem and similarly for other examples. However, they clearly still admit multiple consistent sets without a unifying probability distribution, as long as there are no zero-cover situations. For example, this will be the case in the EPRB example for parameter ranges which violate the CHSH inequalities (except perhaps for very special choices of parameters). That is, they admit situations in which classical notions are violated at a statistical level, but there are no outright contradictions from a classical perspective.
Hence if viewed as a set selection principle, the requirement of a unifying probability is clearly stronger (and so more restrictive) than the requirements of Ordered Consistent Sets, or Preclusive Consistent Sets.

\section{Persistent Classicality}

Another significant example of multiple consistent sets first discussed by Dowker and Kent concerns the question of whether or not a system exhibiting quasi-classical behaviour persists in being quasi-classical into the future \cite{DoK}. Consider a system characterized by alternatives $a_1, a_2, \cdots a_n$ at times $t_1, t_2, \cdots t_n$, which correspond to quasi-classical variables, such as coarse-grained positions. We suppose that their histories are consistent so their probabilities
$p(a_1, a_2, \cdots a_n)$ are well-defined. If these probabilities are strongly peaked around the classical equations of motion (and perhaps a few other reasonable properties hold),  we would say that the histories describe quasi-classical behaviour.

Consider now how these histories may be fine-grained to specify their behaviour to the future of $t_n$, at times
$t_{n+1}, t_{n+2} \cdots t_N $, say. We could consider future alternatives consisting of the same quasi-classical variables, $a_{n+1}, a_{n+2} \cdots a_N$. If these extended histories are consistent their probabilities
$p(a_1, \cdots a_n; a_{n+1} \cdots a_N)$ would be well-defined and quasi-classical behaviour persists to the future.
However, one could also fine-grain to the future using  completely different variables, with alternatives $b_{n+1}, b_{n+2} \cdots b_N$, which could refer to non-classical features of the system. If these histories are consistent we get another well-defined set of probabilities,
$ p(a_1, \cdots a_n; b_{n+1}\cdots b_N)$. These two sets of consistent histories will be incompatible in general, but the second set could exhibit behaviour very far from quasi-classical.

Dowker and Kent showed, on general grounds, that it is possible to construct incompatible consistent sets of this type.
They did not give an explicit example, but these are presumably not hard to find.
They argue that this sort of example makes it difficult to claim that the CH approach predicts emergent classicality since there is no principle favouring either one of these consistent sets. Proponents of the CH approach typically respond by saying that the CH approach makes probabilistic predictions for given sets of histories but remains silent on the issue of whether one set, or the other, or both sets are realized, in any sense. Nevertheless, this particular feature of the CH approach has been a particular source of criticism (see for example Ref.\cite{Smo}).

The approach of the present paper offers an alternative view on this example. As in the previous examples, we ask if there is any unifying probability of the form $ p(a_1, \cdots a_n; a_{n+1}, b_{n+1}, \cdots  a_N, b_N)$, defined on a suitably chosen sample space,
which matches
the two probabilities above obtained from the CH approach. If such a probability exists, it is then consistent to assert that the alternatives in both sets of histories take definite values. If there is no such probability, we cannot make this assertion. Without a more specific example it is difficult to say much more here. However, on the basis of the examples seen previously, we can say that there will be at least some cases in which a unifying probability exists and it is reasonable to talk about both types of future histories as if they both ``happen''. 
Furthermore, we can also say that in the cases where no unifying probability exists, the existence of very different consistent sets is simply a measure of quantumness, and the fact that it is not possible to say in classical terms ``what happens'',
is no more surprising than, for example, the difficulty of saying what happens in situations where the Bell inequalities are violated.

To be clear, this 
is by no means a resolution of the issue in the sense sought by Dowker and Kent, who looked for a principle which would favour certain types of consistent sets over others. It is simply the observation that this disconcerting feature is a reflection of quantumness, so would be a property of any approach to quantum theory, not just the CH approach.

\section{Relationship to the Classification of Quasi-Probabilities}

The approach of this paper -- the idea of finding a unifying probability matching a given set of marginals -- has a clear relationship to recent work on the classification of quasi-probabilities \cite{HaYe}.
In that work it was noted that when quasi-probabilities crop up in quantum mechanics, these are sometimes due to genuinely quantum-mechanical phenomena, but they can also arise in essentially classical situations where there is in fact a genuine probability distribution describing the situation but standard approaches do not automatically reveal it. Hence one needs a way  to distinguish between these two situations, and to construct the probability distribution where it exists.

The approach is as follows. Suppose one is given a quasi-probability $q(a_1,a_2, \cdots a_n)$, for example Eq.(\ref{2.17}). Any quasi-probability will always have a set of marginals which are non-negative. For example, in the case of Eq.(\ref{2.17}) the single time quasi-probabilities obtained by summing out $n-1$ alternatives are non-negative. Suppose that we determine that largest set of non-negative marginals, each obtained by coarse-graining the quasi-probability. Since they are non-negative, they may then be regarded as genuine probabilities for this coarse-grained set of quantities.

Given these marginals one can now ask is there a genuine probability $p(a_1,a_2, \cdots a_n)$ which matches the set of marginals. If there is, the underlying quasi-probability is called ``viable''. If not, it is ``non-viable''. In simple examples, Bell and CHSH inequalities may be used to determine where this probability exists and the difference between non-viable and viable is clearly the difference between a genuinely quantum situation not describable in classical terms and a classical situation for which a probability is not easily found by standard approaches. In simple terms, the marginals of viable quasi-probabilities can be used as if they were marginals of a true probability, and hence without contradiction, whereas
the marginals of non-viable quasi-probabilities cannot be used in this way.

One can now see the relationship between this classification and the classification of multiple consistent sets described in this paper. As noted earlier, given a family of incompatible consistent sets of histories, there is always a quasi-probability, namely Eq.(\ref{2.17}) which has positive marginals matching the probabilities  of each consistent set. Hence the question of the existence or not of a unifying probability coincides precisely with the definitions of viable and non-viable quasi-probabilities.

\section{Summary and Conclusion}

The consistent histories approach has proved to be a very valuable tool for extending the Copenhagen interpretation, understanding the classical limit and delineating the degree to which classical logic may be applied to quantum-mechanical situations. The existence of multiple consistent sets adds subtleties to the interpretation of the approach but the single framework rule provides a clear limitation on what logical deductions can be made.

The consistent histories approach in its standard presentation entails a specific formula for probabilities, Eq.(\ref{prob}), together with specific conditions, namely decoherence or consistency, under which these probabilities are well-defined. The essence of the approach to multiple consistent sets described here is to take a step outside the conventional consistent histories framework and note that decoherence and consistency are part of a larger hierarchy of classicality conditions which includes the weaker condition of linear positivity, and most importantly and weaker still, the technique of finding a unifying probability for a given set of marginal probabilities.
In particular, the present work was based on the simple observation that, if one relaxes focus on the usual probability formula Eq.(\ref{prob}) and associated sample space, then 
some incompatible consistent sets do in fact possess a unifying probability and it is then consistent to assert that some logical deductions can be combined across different consistent sets.
This led to the proposal of an extended single framework rule, allowing a wider set of logical deductions to be made as long as a unifying probability exists. In some examples, this {\it partially} alleviates some of the ontological questions surround multiple consistent sets, i.e. the question of the extent to which one can assign definite values to quantities in different consistent sets.
Of course, it remains true that it is not possible {\it in general} to assign definite values to the alternatives describing incompatible consistent sets, but the proposal put forwards here indicates that it is possible in more situations than previously suspected.

Furthermore, the existence or not of a unifying probability provides a natural definition in the CH approach of the classical-quantum boundary which coincides in a number of examples with intuitive notions and also with other commonly-used (but very weak) classicality measures, such as the Bell, CHSH or Leggett-Garg inequalities, or non-negative Wigner function. In particular, quantumness is seen to be the absence of a unifying probability for certain consistent sets of interest.
This particular issue does not appear to have been addressed previously in the CH approach, which has instead been very focused on the emergence of classical behaviour in the asymptotic regime of histories of very coarse-grained variables exhibiting negligible interference.

Numerous examples of situations both with and without a unifying probability were given. In all cases there was clear accord with intuitive notions of classical or quantum. The proposed classification of consistent histories was compared with earlier (although differently motivated) set selection principles and found to be more restrictive. This work also bears a close relationship with a recently proposed classification of quasi-probabilities and this connection was discussed.

\section{Acknowledgements}

I am grateful to David Craig, Fay Dowker, James Hartle, Adrian Kent, Petros Wallden and James Yearsley for useful conversations. I would also like to thank an anonymous referee for constructive criticisms.



\bibliography{apssamp}

\end{document}